\documentclass[twocolumn,aps,prc,superscriptaddress,showpacs,floatfix]{revtex4}
\usepackage{graphicx}

\begin{document}

\title{Charm elliptic flow at RHIC}
\author{Bin Zhang}
\affiliation{Department of Chemistry and Physics, P.O. Box 419,
Arkansas State University, State University, Arkansas 72467-0419}
\author{Lie-Wen Chen}
\affiliation{Institute of Theoretical Physics, Shanghai Jiao Tong 
University, Shanghai 200240, China}
\author{C. M. Ko}
\affiliation{Cyclotron Institute and Physics Department, Texas A\&M 
University, College Station, Texas 77843-3366}
\date{\today }

\begin{abstract}
Charm elliptic flow in heavy ion collisions at the Relativistic Heavy Ion
Collider (RHIC) is studied in a multiphase transport model. Assuming
that the cross section for charm quark scattering with other light quarks   
is the same as that between light quarks, we find that both charm and 
light quark elliptic flows are sensitive to the value of the cross section.
Compared to that of light quarks, the elliptic flow of charm quarks
is smaller at low transverse momentum but approaches comparable values 
at high transverse momentum. Similar features are seen in the 
elliptic flow of charmed mesons as well as that of the electrons from 
their semileptonic decays when the charmed mesons are produced from
quark coalescence during hadronization of the partonic matter. 
To describe the large electron elliptic flow observed in available
experimental data requires a charm quark scattering cross section 
that is much larger than that given by the perturbative QCD. 
\end{abstract}

\maketitle

\section{Introduction}

Because of their large masses, heavy quarks are produced very early
in ultra-relativistic heavy ion collisions through hard collisions
between nucleons. Their initial momentum spectra can thus be described 
by the perturbative quantum chromodynamics (pQCD). How their final
spectra deviate from the initial ones depends on their interactions in 
the initial partonic matter and the mechanism that converts them to
hadrons as well as their subsequent interactions in the hadronic
matter. Using two extreme scenarios for the charmed meson spectrum,
i.e., pQCD without final-state interactions and completely thermalized 
hydrodynamics with transverse flow velocity field, it was, however,
found in Ref.\cite{Batsouli:2002qf} that the transverse momentum spectra of 
the electrons from their decay are both consistent with that measured 
in Au+Au collisions at \textrm{RHIC} \cite{Adcox:2002cg}. The two scenarios 
give, on the other hand, very different elliptic flows for charmed 
mesons as well as their decay electrons, when the hadronization of 
charm quarks is modeled by coalescence with light quarks
\cite{Greco:2003vf}.  Study of heavy flavor particle elliptic flow in 
ultra-relativistic heavy ion collisions thus provides the possibility 
of probing the interactions of heavy quarks in the partonic matter 
formed in these collisions and their hadronization mechanism.

The study in Ref.\cite{Greco:2003vf} is schematic as the charm quark 
elliptic flow in the scenario of complete thermalization is assumed 
to be the same as the fitted light quark elliptic flow from
experiments. To understand quantitatively the effect 
of final-state interactions on the charmed meson spectrum and elliptic 
flow, we use in the present study a multiphase transport (AMPT)
model that includes both initial partonic and final hadronic 
interactions as well as the transition between these two phases of 
matter. Assuming that the cross section for charm quark scattering 
with light quarks is the same as that between light quarks, we find 
that the charm quark elliptic flow is sensitive to the value of the 
cross section and exhibits a transverse momentum dependence that is 
very different from that of light quarks.  While the elliptic flow of 
charm quarks at low transverse momentum is much smaller than that 
of light quarks as expected from the mass ordering of particle
elliptic flows predicted in the hydrodynamic model, the two become 
comparable at high transverse momentum. For final charmed hadrons and 
their decay electrons, their elliptic flows are found to follow
closely that of charm quarks and thus carry information on the charm quark
interactions in the initial partonic matter. Our results are
similar to that found in Ref.\cite{Molnar:2004ph} based on  
Molnar's Parton Cascade (MPC) model. On the other hand, the value of
the charmed meson elliptic flow from our study is much larger than
that found in Ref.\cite{Bratkovskaya:2004ec} based on the
Hadron-String Dynamics (HSD) model, in which the initial dense matter
consists of strings and hadrons instead of partons as in the AMPT 
and MPC models. The thermalization and elliptic flows of charm quarks
in a hydrodynamically expanding quark-gluon plasma have also been studied 
in a Langevin model, and they are found to be sensitive to the 
charm quark diffusion coefficient \cite{Moore:2004tg}.  We note that 
preliminary results from the present study have been reported in 
Ref.\cite{Chen:2004cx}. 

This paper is organized as follows. In Sect. \ref{ampt}, we give a
brief review of the AMPT model and then extend it to include the 
dynamics of charm particles. Results from the AMPT model on 
both the transverse momentum spectra and elliptic flows of charm 
particles are presented in Sect. \ref{results}. Finally, a summary is
given in Sect. \ref{summary}. 

\section{the AMPT model}\label{ampt}

\subsection{A brief review}

The \textrm{AMPT} model 
\cite{Zhang:2000bd,Lin:2001cx,zhang,ko,pal,Lin:2004en} is a hybrid
model that uses minijet partons from hard processes and strings from soft
processes in the heavy ion jet interaction generator (\textrm{HIJING}) model 
\cite{Wang:1991ht} as the initial conditions for modeling heavy-ion
collisions at ultra-relativistic energies. Since the initial energy density
in Au+Au collisions at \textrm{RHIC} is much larger than the critical
energy density at which the hadronic matter to quark-gluon plasma transition
would occur \cite{zhang,Lin:2004en,Kharzeev:2001ph}, 
we use the version which allows the melting of initial excited strings 
into partons \cite{Lin:2001zk}. In this version, hadrons that would
have been produced from the HIJING model are converted to valence 
quarks and/or antiquarks. Interactions among these partons are 
described by Zhang's parton cascade (\textrm{ZPC}) model 
\cite{Zhang:1997ej}. At present, this model includes only 
parton-parton elastic scatterings with an in-medium cross section 
given by the perturbative QCD, i.e., 
\begin{equation}\label{cross}
\frac{d\sigma _{p}}{dt}=\frac{9\pi \alpha _{s}^{2}}{2}\left( 1+{\frac{{\mu
^{2}}}{s}}\right) \frac{1}{(t-\mu ^{2})^{2}},  \label{crscp}
\end{equation}%
where $\alpha _{s}=0.47$ is the strong coupling constant, and $s$ and $t$
are the usual Mandelstam variables. The effective screening mass $\mu $
depends on the temperature and density of the partonic matter but is taken
as a parameter in \textrm{ZPC} for fixing the magnitude and angular
distribution of parton scattering cross section. Since no inelastic
scatterings are included in the ZPC model, only quarks and antiquarks 
from the melted strings are present in the partonic matter.
The species independence of the above cross section compensates, 
however, for the absence of gluons in the early stage.

After partons stop scattering, they are converted to hadrons using a 
coordinate space quark coalescence model, i.e., two nearest quark and 
antiquark are combined into mesons and three nearest quarks or
antiquarks are combined into baryons or anti-baryons that are closest
to the invariant masses of these parton combinations. This coalescence
model is somewhat different from the ones that are based on the 
overlap of the hadron quark wave functions with the quark distribution 
functions in the partonic matter and used extensively for studying 
the production of hadrons with intermediate transverse momenta 
\cite{greco,hwa,fries}. The final-state scatterings of produced hadrons 
in the AMPT model are described by a relativistic transport
(\textrm{ART}) model \cite{Li:1995pr}. 

Using parton scattering cross sections of $6$-$10$ \textrm{mb}, 
the \textrm{AMPT} model with string melting is able to reproduce the 
centrality and transverse momentum (below $2$ \textrm{GeV}$/c$)
dependence of hadron elliptic flows \cite{Lin:2001zk} and 
higher-order anisotropic flows \cite{Chen:2004dv} as well as the pion 
interferometry \cite{LinHBT02} measured in Au+Au collisions at $\sqrt{s}=130$ 
\textrm{AGeV} at \textrm{RHIC} \cite{Ackermann:2000tr,adams,STARhbt01}. 
It has also been used for studying the kaon interferometry \cite{lin} 
in these collisions as well as many other observables at 
$\sqrt{s}=200$ AGeV \cite{Lin:2004en,Chen:2004vh}.

\subsection{Including charm quarks and charmed mesons}

To include charm particles in the AMPT model, we first generate 
the initial charm quark distributions using information
obtained by the STAR collaboration from d+Au collisions at 
$\sqrt{s_{NN}}=200$ GeV \cite{Adams:2004fc,Tai:2004bf,Ruan:2004if}.
Since final-state interactions are negligible in this collision, 
the measured $D$ meson spectrum is related via the binary collision
scaling to that from the nucleon-nucleon collision at same energy. 
In panel (a) of Fig.~\ref{dndpt1}, we show the extracted transverse
momentum spectrum of $D^0$ mesons in the rapidity interval $|y|\le 1$ 
that are reconstructed from either the $K\pi$ invariant mass (solid 
squares) or the $K\pi\rho$ invariant mass (open circles). Following 
Refs.\cite{Adams:2004fc,Tai:2004bf,Ruan:2004if,Bratkovskaya:2004ec},  
we parameterize this spectrum by a power law,  
\begin{equation}\label{charm}
\frac{d^2N}{2\pi p_Tdp_Tdy} \propto 
\left(1+\frac{p_T}{3.25(\mathrm{GeV}/c)}\right)^{-8.0},
\end{equation}
as shown by the solid curve in Fig.~\ref{dndpt1}(a). The positions of
these charmed mesons are then distributed according to those of
initial binary nucleon-nucleon collisions. 

As for light hadrons produced from string fragmentation in the string 
melting scenario of the AMPT model, we dissociate these \textsl{D} 
mesons into charm and anti-charm quarks according to their valence 
structures to obtain the initial charm quark distributions. The charm 
quarks then propagate along straight-line trajectories from the
positions where they are produced in nucleon-nucleon collisions for a 
duration given by a formation time that is taken to be the inverse of
the $D$ meson transverse momentum.

Since the cross section for charmed meson production in a nucleon-nucleon 
collision extracted from the STAR d+Au data is about 1.3 mb and 
is about a factor of 30 smaller than the nucleon-nucleon 
inelastic cross section ($\sim 40$ mb), the number of charm particles 
produced in relativistic heavy ion collisions is small. To
increase their statistics in the Monte Carlo treatment of the AMPT 
model, we apply the perturbative method that have been extensively 
used in transport models for studying rare particle production in 
heavy ion collisions at lower energies \cite{Randrup:1980qd,Fang:1993ns}. 
Specifically, we scale the charm production cross section by a factor 
that equals the ratio of the nucleon-nucleon inelastic cross section
to that for charm production. Each charm particle thus
carries a probability that equals the inverse of this scale factor. As
experimental data indicate that charm production in heavy ion 
collisions at RHIC is consistent with the binary collision scaling, 
we only allow charm quarks to be produced in initial nucleon-nucleon 
collisions. Rescattering of the charm quarks in the initial 
partonic matter and that of resulting charmed mesons in later hadronic 
matter is, however, included through the AMPT model. Although momenta 
of these charm particles are modified after scattering with other 
more abundant particles, their effects on the latter are neglected 
due to the smaller number of charm particles in each event. 

Both the bulk dynamics of initial partonic matter and the time 
evolution of charm quarks are then described by the ZPC model
using current quark masses, i.e., $9.9$ \textrm{MeV} for \textsl{d} 
quark, $5.6$ \textrm{MeV} for \textsl{u} quark, $199$
\textrm{MeV} for \textsl{s} quark, and $1.35$ \textrm{GeV} for 
\textsl{c} quark. In this study, we use two values for the 
cross section: $\sigma_p=3$ mb estimated from the perturbative QCD 
and $\sigma_p=10$ mb for a strongly interacting QCD. 
Charm rescattering cross sections have been studied before in 
\cite{Combridge:1978kx,Svetitsky:1987gq}, and the 3 mb cross section 
gives an upper bound for the corresponding energy dependent charm 
rescattering cross sections when singularities are regulated by the 
screening mass. We will use, however, the energy independent cross 
section to study whether experimental data require a cross section
much larger than the perturbative one. The differential cross 
section then follows Eq.(\ref{cross}) with $s$ replaced by the maximum
squared momentum transfer for charm rescatterings. As for the
hadronization of light quarks, charm quarks are converted to hadrons 
using the coordinate space quark coalescence model described in the 
previous section. Because of the perturbative method used for treating 
charm quarks, charmed mesons acquire same probabilities as those of 
their parent charm quarks. Although the scattering cross sections of
$D$ mesons with other hadrons have been studied in an effective hadronic 
model \cite{lin1}, we use in the present study instead the same cross 
section as that for parton-parton scattering. We have found that
hadronic scattering does not affect much the final charmed meson 
spectrum and elliptic flow in heavy ion collisions at RHIC. To compare
with measured electrons from charmed meson decays, we include $D$
mesons that are both directly produced as well as from the decay of 
$D^*$ mesons at freeze out. 

\section{results}\label{results}

\subsection{Charm transverse momentum spectra}

\begin{figure}[th]
\includegraphics[scale=0.5]{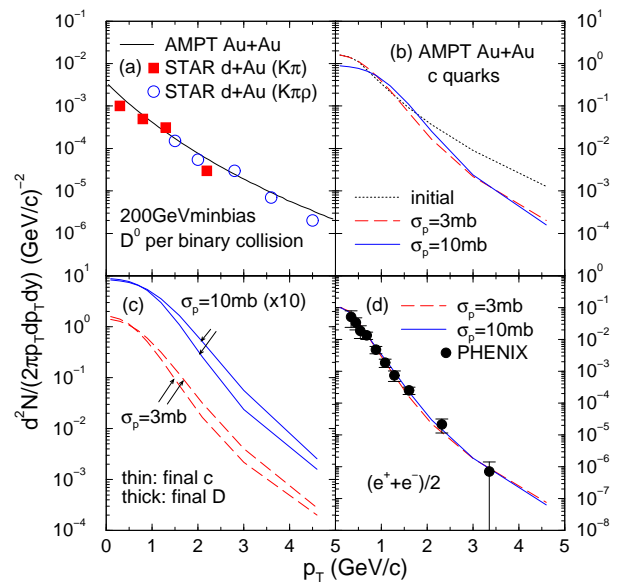}
\caption{(Color online) Transverse momentum distributions at 
midrapidity in minimum bias Au+Au collisions at $\sqrt{s_{NN}}=200$ GeV: (a) 
$D^0$ mesons per binary nucleon-nucleon collision extracted from d+Au
collisions in the STAR experiment \cite{Tai:2004bf} with solid squares 
and open circles denoting data and the solid curve from the parameterization 
(Eq.(\ref{charm})); (b) Initial (dotted curve) and final (dashed curve
for $\sigma_p=3$ mb and solid curve for $\sigma_p=10$ mb) charm quarks 
from AMPT; (c) Final charm quarks (thin dashed and solid curves) and 
$D$ mesons (thick dashed and solid curves); and (d) Electrons from $D$ 
meson decay from AMPT (dashed and solid curves) and the PHENIX
experiment \cite{Adler:2004ta} (solid circles).}
\label{dndpt1}
\end{figure}

The results from the AMPT model on the transverse momentum
distributions of charm particles at midrapidity from minimum bias Au+Au 
collisions at $\sqrt{s_{NN}}=200$ GeV are shown in Fig.~\ref{dndpt1}. 
Panel (b) shows the initial charm quark
transverse momentum spectrum (dotted curve) together with final ones 
obtained using charm quark scattering cross sections of 3 mb 
(dashed curve) and 10 mb (solid curve). The initial charm quark 
transverse momentum spectrum is softer than that of initial $D^0$
mesons shown in panel (a), and this is due to the effective quenching
introduced by converting $D$ mesons into their valence quarks.  
Partonic scattering leads to further quenching of charm quark 
spectrum, with the effect increasing with increasing charm quark
scattering cross section. The quenching effect is, however, reversed
when charm quarks and light quarks are recombined into $D$ mesons 
via the coalescence model as shown in Fig.~\ref{dndpt1}(c), where the 
charm quark spectra at the end of partonic phase (thin curves) and the $D$ 
meson spectra at the end of hadronic phase (thick curves) are
compared. We note that the final $D$ meson number in midrapidity
is about 20\% larger than that of charm quarks in the same rapidity, 
and they are produced from charm quarks at larger rapidity as a result 
of the coordinate space coalescence. 

Although the $D$ meson spectrum in Au+Au collisions has not been 
measured at RHIC, that of the electrons from $D$ meson semi-leptonic
decay is available from the PHENIX collaboration \cite{Adler:2004ta}
as shown by solid circles in panel (d) of Fig.~\ref{dndpt1}. They are 
very well described by the AMPT model with both charm quark scattering 
cross sections of 3 mb (dashed curve) and 10 mb (solid curve). 
Similar to that found in Ref.\cite{Batsouli:2002qf}, these
experimental data thus cannot distinguish the dynamics of charm quarks
in the partonic matter. This is not surprising as the decay process 
significantly softens the transverse momentum spectrum of electrons 
with respect to that of $D$ mesons. We note that in this comparison
the charm production cross section in a nucleon-nucleon collision is 
taken to be 600$\mu$b in order to be consistent with the 
PHENIX measurement.

\subsection{Charm elliptic flow}

We have also studied the elliptic flows of charm quarks and $D$ mesons, 
i.e., the second Fourier coefficient $v_2$ in the decomposition of 
their transverse momentum distributions with respect to the azimuthal 
angle $\phi$ in the reaction plane \cite{Posk98}, 
\begin{equation}
E\frac{d^{3}N}{dp^{3}}=\frac{1}{2\pi }\frac{dN}{p_{T}dp_{T}dy}%
\left[1+\sum_{n=1}^{\infty }2v_{n}(p_{T},y)\cos (n\phi )\right].  
\label{dndphi}
\end{equation}%
Because of the symmetry $\phi \leftrightarrow -\phi $ in the collision
geometry, no sine terms appear in the above expansion. The anisotropic flows 
$v_{n}$ generally depend on the particle transverse momentum and rapidity,
and for a given rapidity the anisotropic flows at transverse momentum $p_{T}$
can be evaluated from 
\begin{equation}
v_{n}(p_{T})=\left\langle \cos (n\phi )\right\rangle,  
\label{vn1}
\end{equation}%
where $\left\langle \cdot \cdot \cdot \right\rangle $ denotes average over
the azimuthal distribution of particles with transverse momentum $p_{T}$.
The elliptic flow $v_{2}$ can further be expressed in terms of the 
single-particle average:
\begin{equation}
v_{2}(p_{T})=\left\langle \frac{p_{x}^{2}-p_{y}^{2}}{p_{T}^{2}}\right\rangle ,
\label{v8}
\end{equation}%
where $p_{x}$ and $p_{y}$ are, respectively, projections of particle
momentum in and perpendicular to the reaction plane.  It has been
shown that the elliptic flow is sensitive to the early dynamics 
of produced matter in relativistic heavy ion collisions
\cite{Sorge:1996pc,Zhang:1999rs,Kolb:2000sd,Teaney:2001av,molnar,Molnar:2004yh},
and it is thus an especially robust observable for studying the 
interactions of early produced particles like charm quarks in the
partonic matter. 

\begin{figure}[th]
\includegraphics[scale=0.5]{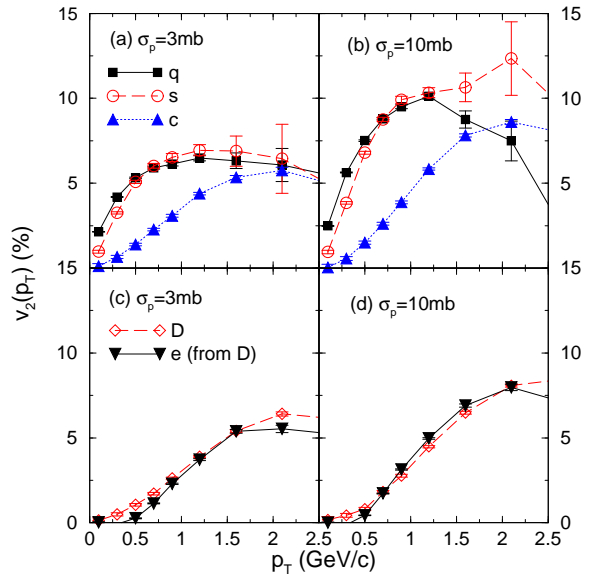}
\caption{(Color online) Elliptic flows of light, strange, and charm 
quarks (upper panels), D mesons, and their decay electrons (lower panels)
in minimum bias Au+Au collisions at $\sqrt{s_{NN}}=200$ GeV
for parton scattering cross sections of $\sigma_p=3$ mb (left panels) 
and $\sigma_p=10$ mb (right panels).}
\label{v2pt2}
\end{figure}

In Fig.~\ref{v2pt2}, we show the transverse momentum dependence of 
elliptic flows of light, strange, and charm quarks as well as those 
of $D$ mesons and electrons from their decays in minimum bias Au+Au 
collisions at $\sqrt{s_{NN}}=200$ GeV.  As shown in the upper panels 
(a) and (b), the $v_{2}$ displays a clear mass dependence, with the 
lighter quark having a larger value of $v_{2}$ at low $p_{T}$ and
reaching the maximum value at a smaller $p_{T}$. This mass effect is 
particularly strong between the heavy charm quark and other light 
quarks. On the other hand, the maximum value of $v_{2}$ is similar for
all quarks, exhibiting thus a very weak quark mass dependence. 
It is, however, sensitive to the parton scattering cross
section, increasing from about $6\%$ to about $9\%$ when the parton 
scattering cross section changes from $\sigma_{p}=3$ \textrm{mb} to 
$10$ \textrm{mb}. The mass dependence of quark $v_{2}$ at lower $p_{T}$
is similar to the mass ordering of particle elliptic flows observed in 
hydrodynamic models. Our results thus imply that with a parton 
scattering cross section larger than $\sigma _{p}=3$ both low $p_T$
charm quarks and light quarks probably approach thermal equilibrium 
in heavy ion collisions at RHIC. 

From the lower panels of Fig.~\ref{v2pt2}, it is seen that the 
\textsl{D} meson elliptic flows (open diamonds) follow closely 
corresponding charm quark elliptic flows shown in the upper panels 
of Fig.~\ref{v2pt2}, although they are slightly shifted to higher
$p_T$ and also have slightly higher values relative to those of charm 
quarks. Also shown in the lower panels of Fig.~\ref{v2pt2} are the 
elliptic flows of electrons from $D$ meson decays (solid triangles), 
and they also follow closely the corresponding \textsl{D} meson ones. 
It is interesting to see that at around $p_T=300$ MeV, the electron 
elliptic flow becomes slightly negative, indicating that on average
these electrons 
are moving perpendicular to their parent \textsl{D} mesons. 
At $p_T$ above 2 GeV, the electron elliptic flow is slightly 
smaller than the $D$ meson elliptic flow. Although the electron $p_T$ 
differential elliptic flow is almost identical to that of \textsl{D} 
mesons, the integrated electron elliptic flow is much smaller than
that of \textsl{D} mesons as the electron transverse spectrum is much 
softer than that of \textsl{D} mesons. We find, e.g.,   
$v_{2e}=0.55\pm 0.01\%\ll v_{2D}=2.73\pm 0.04\%$ and 
$v_{2e}=0.90\pm 0.01\%\ll v_{2D}=3.91\pm 0.04\%$ for parton cross 
sections of 3 and 10 mb, respectively. 

\begin{figure}[th]
\includegraphics[scale=0.6]{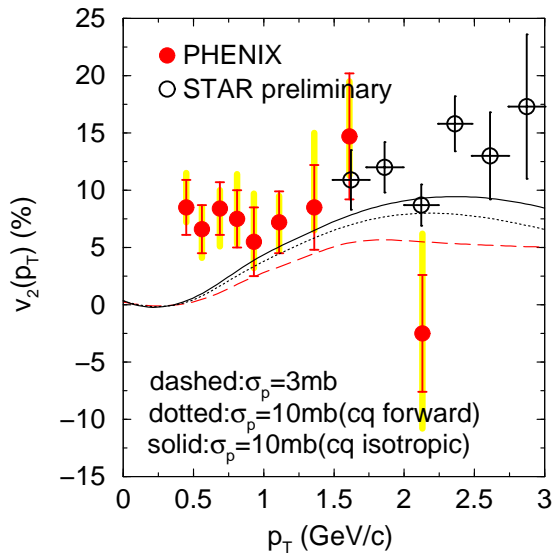}
\caption{(Color online) Transverse momentum dependence of the
elliptic flows of electrons from $D$ meson decays from the AMPT model
for different parton scattering cross sections of 3 (dashed curve) and
10 (dotted curve) mb with angular distribution given by
Eq.(\ref{cross}) as well as 10 mb with isotropic angular
distribution (solid curve) from minimum-bias Au+Au collisions at
$\sqrt{s_{NN}}=200$ GeV. Available experimental data 
are from the PHENIX \cite{Adler:2005ab} (solid circles) and 
STAR \cite{Laue:2004tf} (open circles) collaborations.}
\label{eflow3}
\end{figure}

In Fig.~\ref{eflow3}, we compare the elliptic flow of electrons from
$D$ meson decay obtained from the AMPT model with available experimental 
data from the PHENIX \cite{Adler:2005ab} (solid circles) and STAR 
\cite{Laue:2004tf} (open circles) collaborations. The theoretical
results are shown by the dashed and dotted curves for charm scattering 
cross sections of 3 and 10 mb, respectively.  Compared with the 
experimental data, the calculated charm flow has a similar magnitude 
at $1<p_T<1.5$ Gev/c but is somewhat smaller at other transverse momenta. 

In Ref.\cite{Greco:2003vf}, the elliptic flow of electrons
from $D$ meson decay has also been studied in the coalescence model
based on charm and light quark elliptic flows that are similar to ours
shown in Fig.~\ref{v2pt2}, and its value at high transverse momentum is
larger and thus closer to the experimental data. The smaller charm
elliptic flow from our study is partially due to the quark spatial 
anisotropy, particularly the negative spatial eccentricity $s_2$,
that is present in the dynamical AMPT model but is neglected 
in Ref.\cite{Greco:2003vf}. As shown in Ref. \cite{Pratt:2004zq}, a
negative quark $s_2$ reduces the elliptic flow of produced hadrons
in a dynamic coalescence model such as the one used in the AMPT model. 

As mentioned in Section \ref{ampt}, the magnitude of parton scattering
cross sections used in the AMPT model is determined by the screening
mass or parameter $\mu$ shown in Eq.(\ref{cross}). Since the larger
cross section $\sigma_p=10$ mb is obtained by using a smaller value 
for $\mu$, its angular distribution becomes more forward peaked than 
in the case of 3 mb, thus reducing the effect of increasing cross section 
on the transport properties of quarks. On the other hand, it has 
recently been suggested that the scattering between charm and light 
quarks might go through a quasi colorless charmed resonance, leading 
to a cross section that is not only large but also isotropic 
\cite{vanHees:2004gq}. To see the effect of such a scattering 
mechanism, we have carried out a calculation with an isotropic
charm-light quark scattering cross section of 10 mb, and the elliptic
flow of electrons from the decay of resulting $D$ mesons is shown by 
the solid curve in Fig.~\ref{eflow3}. It is seen that this indeed 
further enhances the charm elliptic flow but not enough to reproduce
available experimental data. If the charm elliptic flow remains
large when more accurate data are available, then the charm quark 
scattering cross section will be even larger than what have been used
in present study. 
 
We note that a naive or additive quark coalescence model 
\cite{Molnar:2003ff,Kolb:2004gi}, which ignores the effect of momentum
distribution of quarks inside hadrons \cite{Lin:2003jy,greco1}, the 
non-negligible local spatial anisotropy \cite{Pratt:2004zq}, and the
large local directed flow \cite{Molnar:2004rr}, has been sometimes
used for calculating the hadron elliptic flows from those of quarks. 
In this case, the elliptic flows of hadrons are then simple sums of 
the elliptic flows of their constituent quarks. Using this simplified 
quark coalescence model, the elliptic flow of electrons from $D$
mesons produced from the charm and light quark elliptic flows shown 
in Fig.~\ref{v2pt2} for a parton scattering cross section of 10 mb 
is comparable to that seen in available experimental data.  

\section{summary}\label{summary}

We have studied the transverse momentum spectra and elliptic 
flows of charm particles and the electrons from charmed meson decay
in heavy ion collisions at RHIC in the AMPT model by treating the 
screening mass in the in-medium pQCD scattering cross section between 
light quarks as a parameter and assuming that charm quarks have 
similar cross sections. Using an initial charm quark transverse
momentum spectrum obtained from the dissociation of the empirically 
determined charmed meson spectrum in nucleon-nucleon collisions 
at same energy, we find that the final transverse momentum spectrum
is softer than that in nucleon-nucleon collisions as a result
of the quenching effect due to charm quark rescattering in the
partonic matter. Although the quenching effect increases with 
increasing charm quark scattering cross section, the transverse 
momentum spectrum of electrons from resulting $D$ meson decays 
does not depend much on the parton scattering cross sections
as a result of the stronger quenching effect due to the decay process, 
and results using both cross sections of 3 and 10 mb are consistent
with available experimental data from the PHENIX experiment. 

The elliptic flow of 
charm quarks is, however, sensitive to the parton cross section 
as that of light quarks. With respect to the light quark elliptic
flow, the charm quark elliptic flow is smaller at low transverse 
momentum but reaches a similar value at high transverse momentum. With
the coordinate space quark coalescence model for hadronization in 
the AMPT model, the elliptic flow of $D$ mesons is found to follow
that of charm quarks. Compared to measured elliptic flow of electrons 
from $D$ meson decay in Au+Au collisions at $\sqrt{s}=200$
\textrm{AGeV}, the calculated values from the AMPT model, which is 
close to the $D$ meson elliptic flow, are somewhat smaller even for 
an isotropic charm-light quark scattering cross section of 10 mb. 
Such a cross section is much larger than that given by the 
perturbative QCD estimate but is consistent with that based on novel 
resonant heavy-light quark interactions \cite{vanHees:2004gq} inside a 
strongly interacting quark-gluon plasma consisting of quasi-particles
and colored bound states \cite{Shuryak:2003ty}. The
study of charm flow in relativistic heavy ion collisions thus provides a 
promising tool to explore the properties of produced partonic matter 
and its hadronization dynamics. 

\begin{acknowledgments}
We thank Vincenzo Greco, Miklos Gyulassy, Zi-Wei Lin, Scott Pratt, 
Ralf Rapp, and An Tai for useful discussions.  This paper was based on work 
supported by the U.S. National Science Foundation under Grant No.'s 
PHY-0098805 and PHY-0457265 (CMK) and PHY-0140046 (BZ), the Welch Foundation 
under Grant No. A-1358 (CMK), and the National Natural Science 
Foundation of China under Grant No. 10105008 (LWC).
\end{acknowledgments}


\begin{thebibliography}{99}

\bibitem{Batsouli:2002qf}
S.~Batsouli, S.~Kelly, M.~Gyulassy and J.~L.~Nagle,
Phys.\ Lett.\ B {\bf 557}, 26 (2003).

\bibitem{Adcox:2002cg}
K.~Adcox {\it et al.}  [PHENIX Collaboration],
Phys.\ Rev.\ Lett.\  {\bf 88}, 192303 (2002)
;Nucl. Phys. A {\bf 715}, 695 (2003).

\bibitem{Greco:2003vf}
V.~Greco, C.~M.~Ko and R.~Rapp,
Phys.\ Lett.\ B {\bf 595}, 202 (2004).

\bibitem{Molnar:2004ph}
D.~Molnar,
J.\ Phys.\ G {\bf 31}, S421 (2005).

\bibitem{Bratkovskaya:2004ec}
E.~L.~Bratkovskaya, W.~Cassing, H.~Stocker and N.~Xu,
Phys.\ Rev.\ C {\bf 71}, 044901 (2005)

\bibitem{Moore:2004tg}
G.~D.~Moore and D.~Teaney,
arXiv:hep-ph/0412346.

\bibitem{Chen:2004cx}
L.~W.~Chen and C.~M.~Ko,
J.\ Phys.\ G {\bf 31}, S49 (2005).

\bibitem{Zhang:2000bd} {\small B.~Zhang, C. M. Ko, B. A. Li and Z. W. Lin,
Phys.\ Rev.\ C \textbf{61}, 067901 (2000).}

\bibitem{Lin:2001cx} {\small Z.~W.~Lin, S. Pal, C. M. Ko, B. A. Li, and B.
Zhang, Phys.\ Rev.\ C \textbf{64}, 011902 (2001); Nucl.\ Phys.\ \textbf{A698}%
, 375 (2002).}

\bibitem{zhang} {\small B.~Zhang, C. M. Ko, B. A. Li, Z. W. Lin, and B. H.
Sa, Phys.\ Rev.\ C \textbf{62}, 054905 (2000); B. Zhang, C. M. Ko, B. A. Li,
Z. W. Lin, and S. Pal, \textit{ibid.} \textbf{65}, 054909 (2002).}

\bibitem{ko} {\small C.M. Ko, Z.W. Lin, and S. Pal, Heavy Ion Phys. \textbf{%
17}, 219 (2003).}

\bibitem{pal} {\small S. Pal. C. M. Ko, and Z. W. Lin, Nucl. Phys. \textbf{%
A730}, 143 (2004).}

\bibitem{Lin:2004en}
Z.~W.~Lin, C.~M.~Ko, B.~A.~Li, B.~Zhang and S.~Pal,
arXiv:nucl-th/0411110.

\bibitem{Wang:1991ht} {\small X.~N.~Wang and M.~Gyulassy, Phys.\ Rev.\ D 
\textbf{44}, 3501 (1991).}

\bibitem{Kharzeev:2001ph} {\small D.~Kharzeev and M.~Nardi, Phys.\ Lett.\ 
\textbf{B507}, 121 (2001).}

\bibitem{Lin:2001zk} {\small Z.~W.~Lin and C.~M.~Ko, Phys.\ Rev.\ C \textbf{%
65}, 034904 (2002).}

\bibitem{Zhang:1997ej} {\small B.~Zhang, Comput.\ Phys.\ Commun.\ \textbf{109%
}, 193 (1998).}

\bibitem{greco} {\small V. Greco, C.M. Ko, and P. L\'{e}vai, Phys. Rev.
Lett. \textbf{90}, 202302 (2003); Phys. Rev. C \textbf{68}, 034904 (2003).}

\bibitem{hwa} {\small R.C. Hwa and C.B. Yang, Phys. Rev. C \textbf{67},
034902 (2003); 064902 (2003).}

\bibitem{fries} {\small R.J. Fries, B. M\"{u}ller, C. Nonaka, and S.A. Bass,
Phys. Rev. Lett. \textbf{90}, 202303 (2003); Phys. Rev. C \textbf{68},
044902(2003).}

\bibitem{Li:1995pr} {\small B.~A.~Li and C.~M.~Ko, Phys.\ Rev.\ C \textbf{52}%
, 2037 (1995); B.A. Li, A.T. Sustich, B. Zhang, and C.M. Ko, Int. Jour.
Phys. E \textbf{10}, 267-352 (2001).}

\bibitem{Chen:2004dv}
L.~W.~Chen, C.~M.~Ko and Z.~W.~Lin,
Phys.\ Rev.\ C {\bf 69}, 031901 (2004)

\bibitem{LinHBT02} {\small Z.~W.~Lin, C.~M.~Ko and S. Pal, Phys.\ Rev.\ \
Lett.\ \textbf{89}, 152301 (2002).}

\bibitem{Ackermann:2000tr} 
{\small K.~H.~Ackermann \textit{et al.}, [STAR Collaboration], 
Phys.\ Rev.\ Lett.\ \textbf{86}, 402 (2001).}

\bibitem{adams}J. Adams {\it et al.}, [STAR Collaboration],
Phys. Rev. lett. {\bf 92}, 062301 (2004).

\bibitem{STARhbt01} 
{\small C.~Adler \textit{et al.}, [STAR Collaboration], Phys.\
Rev.\ Lett.\ \textbf{87}, 082301 (2001).}

\bibitem{lin} {\small Z. W. Lin and C. M. Ko, J. Phys. G \textbf{30}, S263
(2004).}

\bibitem{Chen:2004vh}
L.~W.~Chen, V.~Greco, C.~M.~Ko and P.~F.~Kolb,
Phys.\ Lett.\ B {\bf 605}, 95 (2005).

\bibitem{Adams:2004fc}
J.~Adams \textit{et al.}  [STAR Collaboration],
Phys.\ Rev.\ Lett.\  {\bf 94}, 062301 (2005).

\bibitem{Tai:2004bf}
A.~Tai  [STAR Collaboration],
J.\ Phys.\ G {\bf 30}, S809 (2004).

\bibitem{Ruan:2004if}
L.~Ruan [STAR Collaboration],
J.\ Phys.\ G {\bf 30}, S1197 (2004).

\bibitem{Randrup:1980qd}
J.~Randrup and C.~M.~Ko,
Nucl.\ Phys.\ A {\bf 343}, 519 (1980).

\bibitem{Fang:1993ns}
X.~S.~Fang, C.~M.~Ko and Y.~M.~Zheng,
Nucl.\ Phys.\ A {\bf 556}, 499 (1993).

\bibitem{Combridge:1978kx}
B.~L.~Combridge,
Nucl.\ Phys.\ B {\bf 151}, 429 (1979).

\bibitem{Svetitsky:1987gq}
B.~Svetitsky,
Phys.\ Rev.\ D {\bf 37}, 2484 (1988).

\bibitem{lin1}Z. W. Lin, C. M. Ko, and B. Zhang, Phys. Rev. C {\bf 61}, 
024904 (2000); Z. W. Lin, T. G. Di, and C. M. Ko, Nucl. Phys. A 
{\bf 689}, 965 (2001). 

\bibitem{Adler:2004ta}
S.~S.~Adler \textit{et al.}  [PHENIX Collaboration],
Phys.\ Rev.\ Lett.\  {\bf 94}, 082301 (2005).

\bibitem{Posk98} {\small A.~M.~Poskanzer and S.~A.~Voloshin, Phys.\ Rev.\ C 
\textbf{58}, 1671 (1998).}

\bibitem{Sorge:1996pc}
H.~Sorge,
Phys.\ Rev.\ Lett.\  {\bf 78}, 2309 (1997).

\bibitem{Zhang:1999rs}
B.~Zhang, M.~Gyulassy and C.~M.~Ko,
Phys.\ Lett.\ B {\bf 455}, 45 (1999).

\bibitem{Kolb:2000sd}
P.~F.~Kolb, J.~Sollfrank and U.~W.~Heinz,
Phys.\ Rev.\ C {\bf 62}, 054909 (2000).

\bibitem{Teaney:2001av}
D.~Teaney, J.~Lauret and E.~V.~Shuryak,
arXiv:nucl-th/0110037.

\bibitem{molnar} {\small D. Molnar and M. Gyulassy, Nucl. Phys. A697, 495
(2002); A703, 893 (2002).}

\bibitem{Molnar:2004yh}
D.~Molnar and P.~Huovinen,
Phys.\ Rev.\ Lett.\  {\bf 94}, 012302 (2005).

\bibitem{Adler:2005ab}
S.~S.~Adler  [PHENIX Collaboration],
arXiv:nucl-ex/0502009.

\bibitem{Laue:2004tf}
F.~Laue  [STAR Collaboration],
J.\ Phys.\ G {\bf 31}, S27 (2005).

\bibitem{Pratt:2004zq}
S.~Pratt and S.~Pal,
Phys.\ Rev.\ C {\bf 71}, 014905 (2005).

\bibitem{vanHees:2004gq}
H.~van Hees and R.~Rapp,
Phys.\ Rev.\ C {\bf 71}, 034907 (2005).

\bibitem{Kolb:2004gi}
P.~F.~Kolb, L.~W.~Chen, V.~Greco and C.~M.~Ko,
Phys.\ Rev.\ C {\bf 69}, 051901 (2004)

\bibitem{Molnar:2003ff}
D.~Molnar and S.~A.~Voloshin,
Phys.\ Rev.\ Lett.\  {\bf 91}, 092301 (2003)

\bibitem{Lin:2003jy}
Z.~W.~Lin and D.~Molnar,
Phys.\ Rev.\ C {\bf 68}, 044901 (2003).

\bibitem{greco1}V. Greco and C. M. Ko, Phys. Rev. C {\bf 70}, 024901 (2004).

\bibitem{Molnar:2004rr}
D.~Molnar,
arXiv:nucl-th/0408044.

\bibitem{Shuryak:2003ty}
E.~V.~Shuryak and I.~Zahed,
Phys.\ Rev.\ C {\bf 70}, 021901 (2004); 
Phys.\ Rev.\ D {\bf 70}, 054507 (2004).

\end{thebibliography}
\end{document}